\documentclass{elsart}

\usepackage{graphicx}
\usepackage{subfigure}
\usepackage{bmpsize}
\usepackage[parfill]{parskip}

\usepackage{amssymb}

\usepackage{multirow}
\usepackage{color}

\begin{document}

\begin{frontmatter}



\title{Light yield determination in large sodium iodide detectors applied in the search for dark matter}


\author[UZ,LSC]{M.A. Oliv{\'a}n},
\author[UZ,LSC]{J.~Amar{\'e}},
\author[UZ,LSC]{S.~Cebri{\'an}},
\author[UZ,LSC,CIEMAT]{C.~Cuesta},
\author[UZ,LSC]{E.~Garc\'{\i}a},
\author[UZ,LSC,UR]{M.~Mart\'{\i}nez},
\author[UZ,LSC]{Y.~Ortigoza},
\author[UZ,LSC]{A.~Ortiz~de~Sol{\'o}rzano},
\author[UZ,LSC,ICMA]{C.~Pobes},
\author[UZ,LSC]{J.~Puimed{\'o}n},
\author[UZ,LSC]{M.L.~Sarsa\thanksref{CA}},
\author[UZ,LSC]{J.A.~Villar}, and
\author[UZ,LSC]{P.~Villar}

\address[UZ]{Laboratorio de F\'{\i}sica Nuclear y Astropart\'{\i}culas, Universidad de Zaragoza, C/~Pedro Cerbuna 12, 50009 Zaragoza, SPAIN}
\address[LSC]{Laboratorio Subterr{\'a}neo de Canfranc, Paseo de los Ayerbe s.n., 22880 Canfranc Estaci{\'o}n, Huesca, SPAIN}
\address[CIEMAT]{Present Address: Centro de Investigaciones Energ{\'e}ticas, Medioambientales y Tecnol{\'o}gicas, CIEMAT, 28040, Madrid, SPAIN}
\address[UR]{Present Address: Universit{\`a} di Roma La Sapienza, Piazzale Aldo Moro 5, 00185 Rome, ITALY}
\address[ICMA]{Present Address: Instituto de Ciencia de Materiales de Arag{\'o}n, CSIC-Universidad de Zaragoza, Zaragoza, SPAIN}
\thanks[CA]{Corresponding author (mlsarsa@unizar.es)}

\begin{abstract}

Application of NaI(Tl) detectors in the search for galactic dark matter particles through their elastic scattering off the target nuclei is well motivated because of the long standing DAMA/LIBRA highly significant positive result on annual modulation, still requiring confirmation. For such a goal, it is mandatory to reach very low threshold in energy (at or below the keV level), very low radioactive background (at a few counts/keV/kg/day), and high detection mass (at or above the 100 kg scale). One of the most relevant technical issues is the optimization of the crystal intrinsic scintillation light yield and the efficiency of the light collecting system for large mass crystals. In the frame of the ANAIS (Annual modulation with NaI Scintillators) dark matter search project large NaI(Tl) crystals from different providers coupled to two photomultiplier tubes (PMTs) have been tested at the Canfranc Underground Laboratory. In this paper we present the estimates of the NaI(Tl) scintillation light collected using full-absorption peaks at very low energy from external and internal sources emitting gammas/electrons, and single-photoelectron events populations selected by using very low energy pulses tails. Outstanding scintillation light collection at the level of 15~photoelectrons/keV can be reported for the final design and provider chosen for ANAIS detectors. Taking into account the Quantum Efficiency of the PMT units used, the intrinsic scintillation light yield in these NaI(Tl) crystals is above 40~photoelectrons/keV for energy depositions in the range from 3 up to 25~keV. This very high light output of ANAIS crystals allows triggering below 1~keV, which is very important in order to increase the sensitivity in the direct detection of dark matter.

\end{abstract}

\end{frontmatter}

\section{Introduction}
\label{intro}
Thallium Activated Sodium Iodide (NaI(Tl)) scintillators have been widely used for radiation detection \cite{Birks, Lecoq} since they were proposed by R. Hofstadter in 1948 \cite{Hofstadter}. Among other remarkable features, the very high intrinsic scintillation light yield has been probably a key factor to widespread their use. Sodium Iodide scintillators have been applied very satisfactorily in fields so different as: nuclear medicine, environmental monitoring, nuclear physics, aerial survey, well logging, homeland security, etc. Energy ranges from 10~keV to several MeV can be covered with state-of-the-art technology, but applications requiring high sensitivity at lower energies have moved to other detection techniques having better energy resolution and/or lacking from the hygroscopic property of NaI, that makes difficult the access of low energy radiation to the sensitive volume. However, these detectors have been successfully applied since the nineties in the direct search for dark matter in the form of WIMPs (Weakly Interacting Massive Particles), that should pervade the galactic halo and because of the very low interaction probability with matter they have, they could reach that sensitive volume independently of the encapsulation or even the placement of the detector in an underground location, as it is usually done in order to reduce the cosmic-ray-induced background. Several experimental efforts using NaI(Tl) detectors applied in the search for dark matter can be found in the literature \cite{JAPAN1,DM32,DAMA,PSD_Gerbier,UKDMS,DAMALIBRA}, as well as several proposals for the next future \cite{ANAIS,KIMS,DM-ice,SABRE,PICO-LON}.

The ANAIS project aims at the installation of a 112.5~kg NaI(Tl) experiment at the Canfranc Underground Laboratory (LSC), in Spain, to study the annual modulation effect expected in the interaction rate of dark matter particles as result of the Earth motion around the Sun \cite{Freese}. In the frame of ANAIS a long effort has been carried out in understanding radioactive backgrounds in NaI(Tl) detectors \cite{ANAISbckgmodel,ANAISbckgmodel2,40K,cosmogenia} and in the  assessment of NaI(Tl) detectors performance \cite{cuarzo,slowscint,filtrado}. For such a purpose, crystals from different providers (BICRON, Saint-Gobain and Alpha Spectra) have been studied. In this work, we will present our estimates of total light collected at the PMTs for low energy depositions produced using external sources, but also internal contaminations homogeneously distributed in the crystal bulk. Total light seen by the light detection system in scintillating detectors is one of the most relevant figures of merit, contributing strongly to the achievable energy threshold, succesful implementation of PMT noise rejection procedures, and energy resolution. It depends on intrinsic scintillation light yield of the scintillating material and light collection efficiency of the detection system, including self-absorption, trapped light by total internal reflection, light losses by the multiple reflections and scattering in the non-sensitive detector surfaces, and Quantum Efficiency (QE) of the PMTs used as light sensors. It can be determined as photoelectrons (p.e.) produced at PMT photocathode and reaching the first dynode per unit of energy deposited by comparing the signal produced by the full-absorption of a given energy in the scintillating material and the signal corresponding to a single p.e.; then, it can be converted into scintillation photons reaching the PMT photocathode by correcting by the corresponding QE of the PMT units used. However, intrinsic scintillation light yield for NaI(Tl) is not easy to deconvolve from the total light collected measurement, as light collection efficiency should be known. Knoll and others \cite{Knoll} suggested the intrinsic scintillation efficiency in NaI(Tl) to be 40,000~photons/MeV, implying a W$_s$ value of 25~eV for the average energy required to produce a scintillation photon. However, there are theoretical estimates of lower W$_s$ value in NaI(Tl), resulting in a maximum scintillation efficiency of about 100,000~photons/MeV \cite{Doke}. The high light collection values reported in this work for very large NaI(Tl) crystals using gamma/electron radiation at very low energy could be an interesting input for such calculations, as far as most of the previous reports and analysis are not reaching such low energy range \cite{Knoll, Doke, Murray} and have been obtained with smaller crystals, for which light propagation and collection is not expected to be a so much relevant issue; moreover, some aspects of the scintillation mechanism in NaI(Tl) crystals are not yet fully understood \cite{JAP,NIM_Mos}. 

The structure of this paper is the following: in section {\ref{set-up}} we describe the ANAIS prototype modules used in the reported measurements and in section {\ref{data-acq}}, the data acquisition system. In section {\ref{analysis}} we describe our analysis procedure to derive the Single Electron Response (SER) and the full-absorption lines considered for the determination of the light collected at each PMT. Finally, in the last section results are presented and conclusions drawn.

\section{Modules Description}
\label{set-up}
\begin {figure}[htb]
\centerline{\subfigure[]{\includegraphics[width=0.4\textwidth]{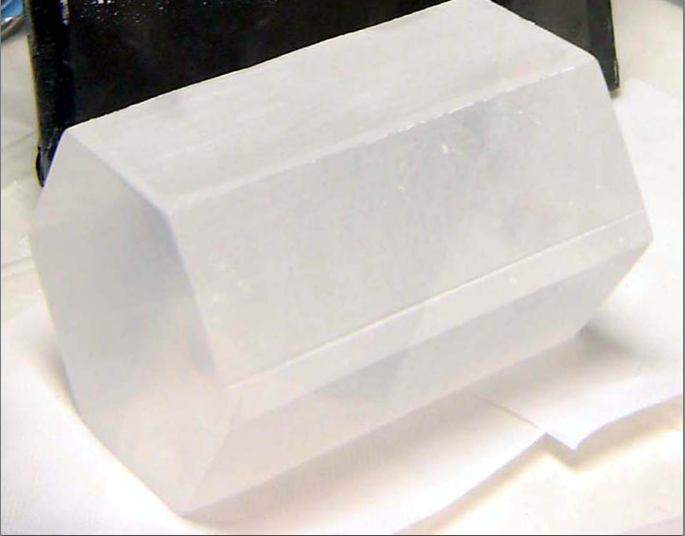}}
\subfigure[] {\includegraphics[width=0.22\textwidth]{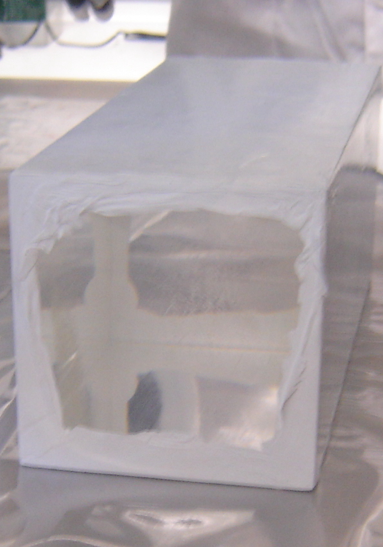}}
\subfigure[]{\includegraphics[width=0.42\textwidth]{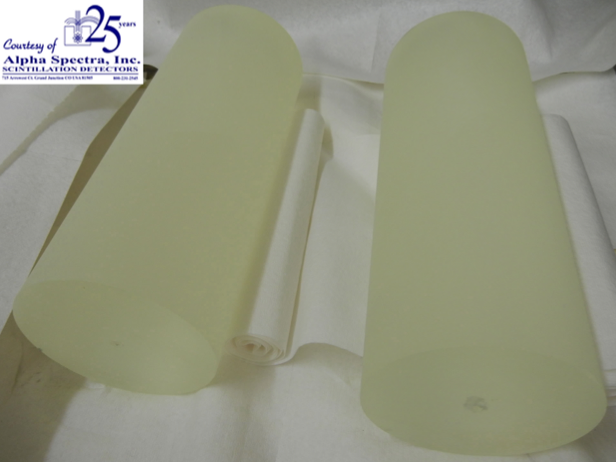}}}
\caption{Different NaI(Tl) crystals studied at the LSC in order to determine their total light yield: (a) 10.7\,kg crystal from BICRON (PIII), (b) 9.7\,kg crystal from Saint Gobain (A0), and (c) 12.5\,kg crystals from Alpha Spectra (D0/D1).}
\label{fig:crystals}
\end{figure}

\begin{figure}[htb]
\begin{center}
\subfigure[]{\includegraphics[width=0.65\textwidth]{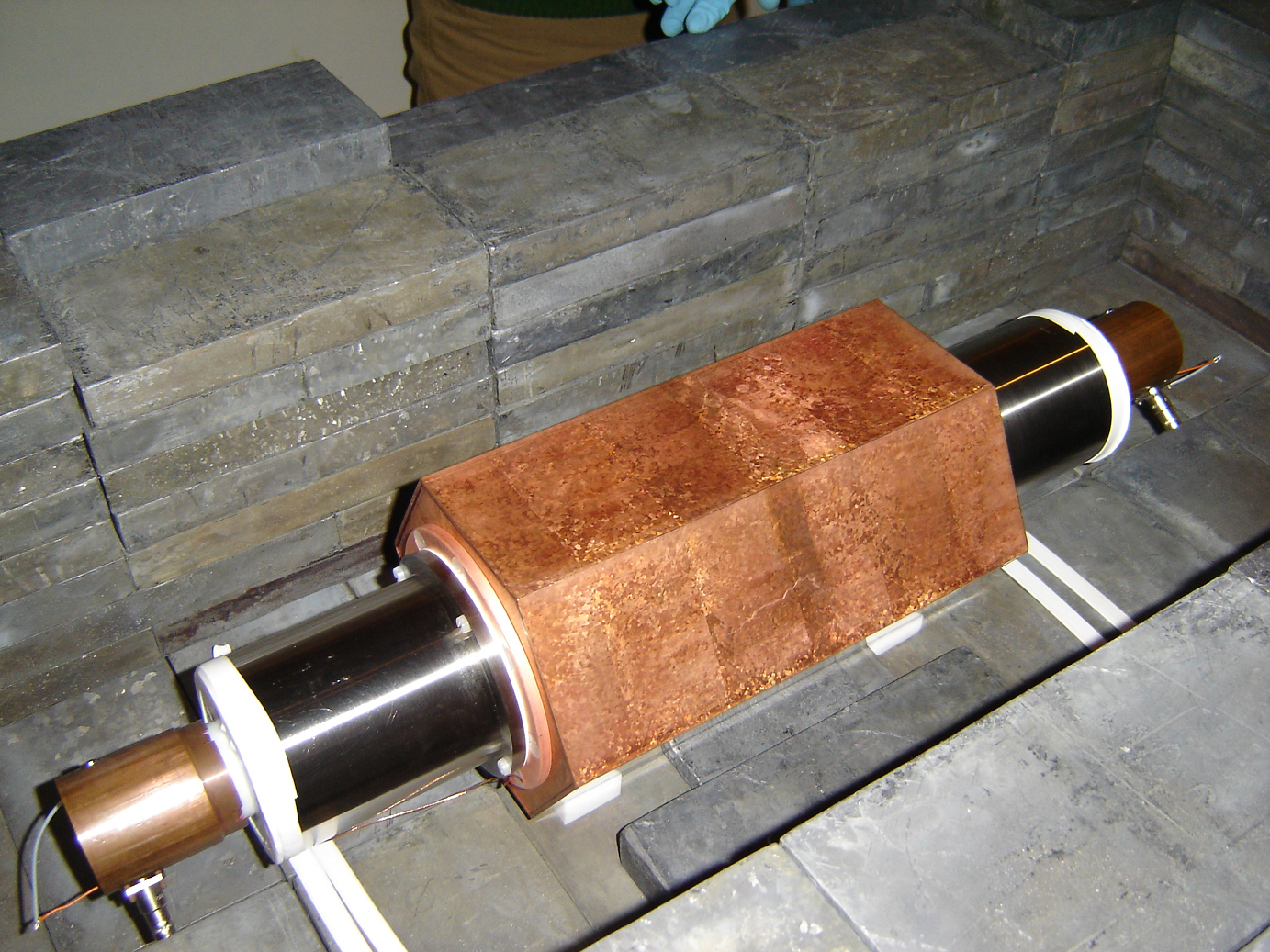}}
\subfigure[]{\includegraphics[width=0.85\textwidth]{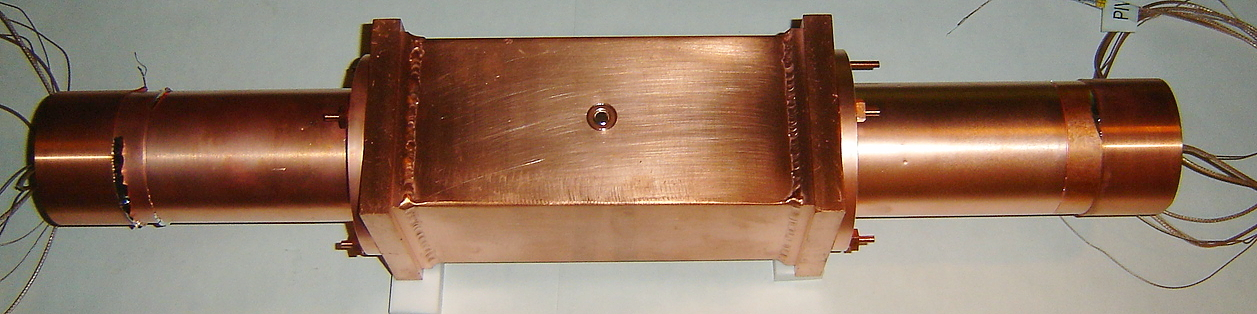}}
\subfigure[]{\includegraphics[width=0.85\textwidth]{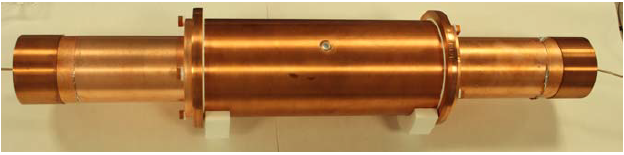}}
\caption{NaI(Tl) built modules used to determine their total light yield: (a)~PIII, (b)~A0, and (c)~AS-D0 module.}\label{fig:detectors}
\end{center}
\end{figure}

In this work, crystals from three different manufacturers, and having different size and shape, have been characterized in terms of the light collected at each PMT and in total at each crystal: 
\begin{itemize}
\item A 10.7~kg hexagonal prism NaI(Tl) crystal (distance between opposite vertices in the hexagonal face 15.94~cm, and 20.32~cm high) disassembled from the original stainless steel encapsulation made by BICRON. It was polished and cleaned at the University of Zaragoza (see Fig.~\ref{fig:crystals}.a). A new 1~mm-thick Oxygen Free High Conductivity (OFHC) copper encapsulation was specifically designed for this prototype, shown in Fig.~\ref{fig:detectors}.a. In the following, we will refer to this detector as PIII. 
\item A 9.6~kg parallelepiped ultrapure NaI(Tl) crystal (4"~x~4"~x~10"), made by Saint-Gobain, was encapsulated at the University of Zaragoza using 1~mm-thick Electrolytic Tough Pitch (ETP) copper, and it is shown in Fig.~\ref{fig:crystals}.b, and Fig.~\ref{fig:detectors}.b. In the following, this module is labeled A0.
\item Several units of 12.5~kg crystals made by Alpha Spectra Inc. (AS) using a low-potassium content NaI powder and encapsulated in Oxygen Free Electrolytic (OFE) copper at AS. They are cylindrical, with diameter of 4.75" and 11.75" length. We include below data from the first four units AS has built for ANAIS-112 experiment, that are labelled as D0, D1, D2, and D3. In Fig.~\ref{fig:crystals}.c and Fig.~\ref{fig:detectors}.c, are shown D0 and D1 crystals, and D0 module, respectively. 
\end{itemize}

All the modules share some common features: 
\begin{itemize}
\item Low background Hamamatsu PMTs of different models have been used for the tests. QE nominal values from manufacturer for the different units used in the reported measurements are given in Table~\ref{tab:PMTs}.
\item An aluminized Mylar window (20~$\mu$m thick and 10~mm diameter) in the middle of one of the lateral faces allows for external calibration of all the modules at very low energies (see Fig.~\ref{fig:detectors}).
\item Tight sealing is done at the level of the two 3" diameter quartz optical windows and two PMTs are coupled to each crystal in a second step:
\begin{itemize}
\item 1.27~cm long natural quartz optical windows were used for PIII module;
\item 1~cm long synthetic quartz optical windows were used for A0 and all the AS modules.
\end{itemize}
\end{itemize}

All data have been taken at the facilities of the LSC, in Spain, under 2450 m.w.e. Detectors, in all the different set-ups considered in this work, were installed in a shielding consisting of 10~cm archaeological lead plus 20~cm low activity lead, all enclosed in a PVC box tightly closed and continuously flushed with nitrogen gas. For the here reported analysis we have used data from calibration runs, but also background data; for the latter, convenient shielding and underground location of the modules are mandatory. Next, we list some details of the different set-ups for which results will be presented in section~\ref{analysis}. 
\begin{itemize}
\item A0 set-ups: A0 module was taking data in different experimental conditions in order to choose the optimum configuration in terms of threshold and background: different PMT models were tested, and light guides effect in light collection and background contributions were studied~\cite{CCuestaTesis}. Data were taken in a long period between April 2009 and August 2012. Data used below correspond to set-ups A0-3 (May - June 2011) using 10~cm long methacrylate light guides, A0-4 (June - December 2011) and A0-5 (March - December 2012), both without light guides. PIII and A0 modules took data in the same set-up (A0-5) at LSC from August until December 2012 in order to precisely measure the $^{40}$K content in the latter. 
\item ANAIS25 set-up: two AS modules (D0 and D1) took data at LSC from December 2012 until March 2015 to determine their internal contaminations and general performance assessment; because of that, several upgrades were implemented, for instance, PMTs in D1 module were replaced. Data shown below correspond to phases I and III of this set-up. 
\item ANAIS37 set-up: a new module (D2) was placed in between D0 and D1. It took data from March 2015 until March 2016 in order to characterize such new module D2. 
\item A37D3 set-up: module D1 was removed and a new module (D3) was placed in between D0 and D2, having similar goals as previous set-up. Data included below correspond to the first operation period, from March until June 2016. 
\end{itemize}

\begin{table}
\begin{center}
\begin{tabular}{| c  c | c |}
\hline
 PMT model & PMT unit reference & QE (\%)     \\
\hline
Ham R11065SEL & BA0086 &  28.68  \\
 & BA0057 & 32.90 \\
\hline
Ham R6956MOD & ZK5902  & 34.9 \\
			         & ZK5908  & 34.4 \\
\hline
Ham R12669SEL2  & FA0018 & 33.9   \\
                     & FA0060 & 36.4    \\
                     & FA0034 & 35.9   \\
                     & FA0090 & 39.0   \\
\hline

\end{tabular}
\end{center}
\vspace{0.2cm}
\caption{Quantum Efficiency of the PMT units used in the determination of the NaI crystals total light yield presented in this work. Provided by manufacturer.}
\label{tab:PMTs}
\vspace{0.5cm}
\end{table}

\section{Electronics and data acquisition}
\label{data-acq}
\begin{figure}[htb]
\begin{center}
\subfigure[]{\includegraphics[width=0.8\textwidth]{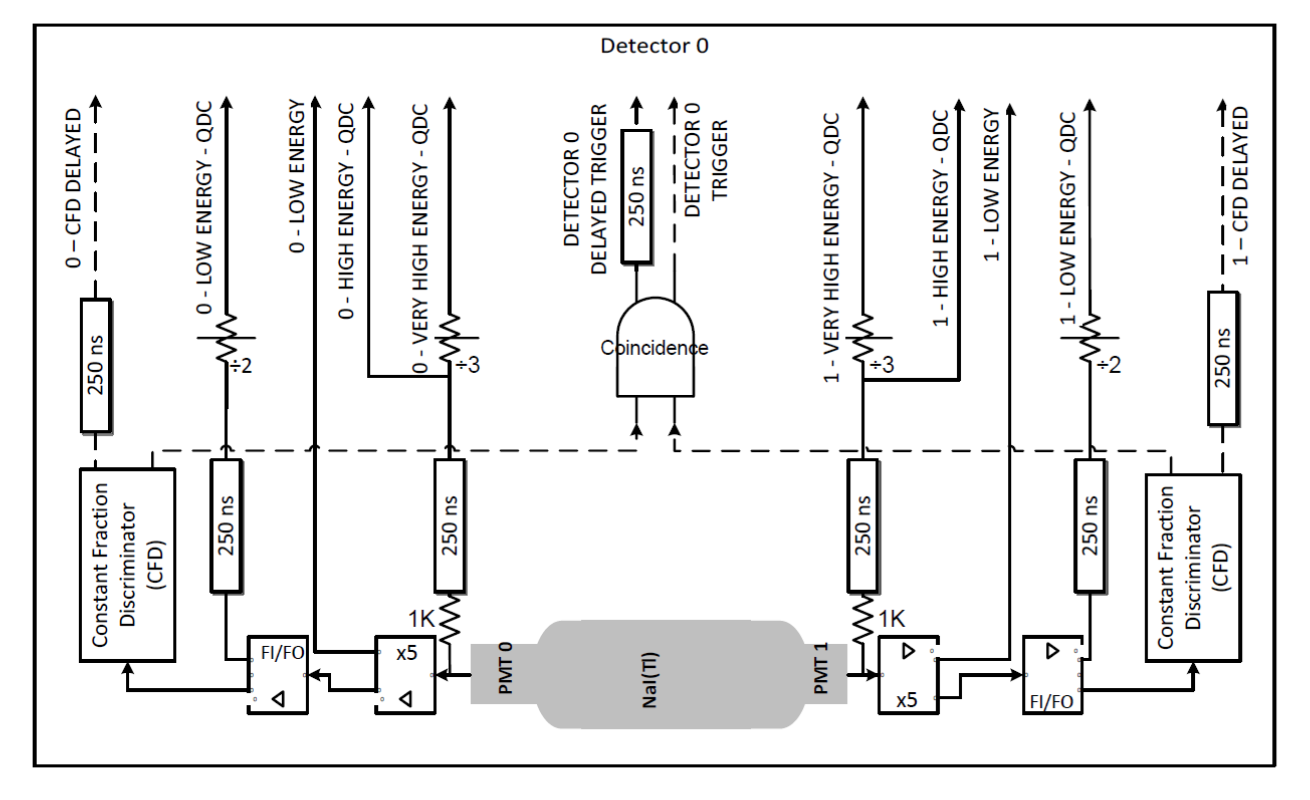}}
\subfigure[]{\includegraphics[width=0.8\textwidth]{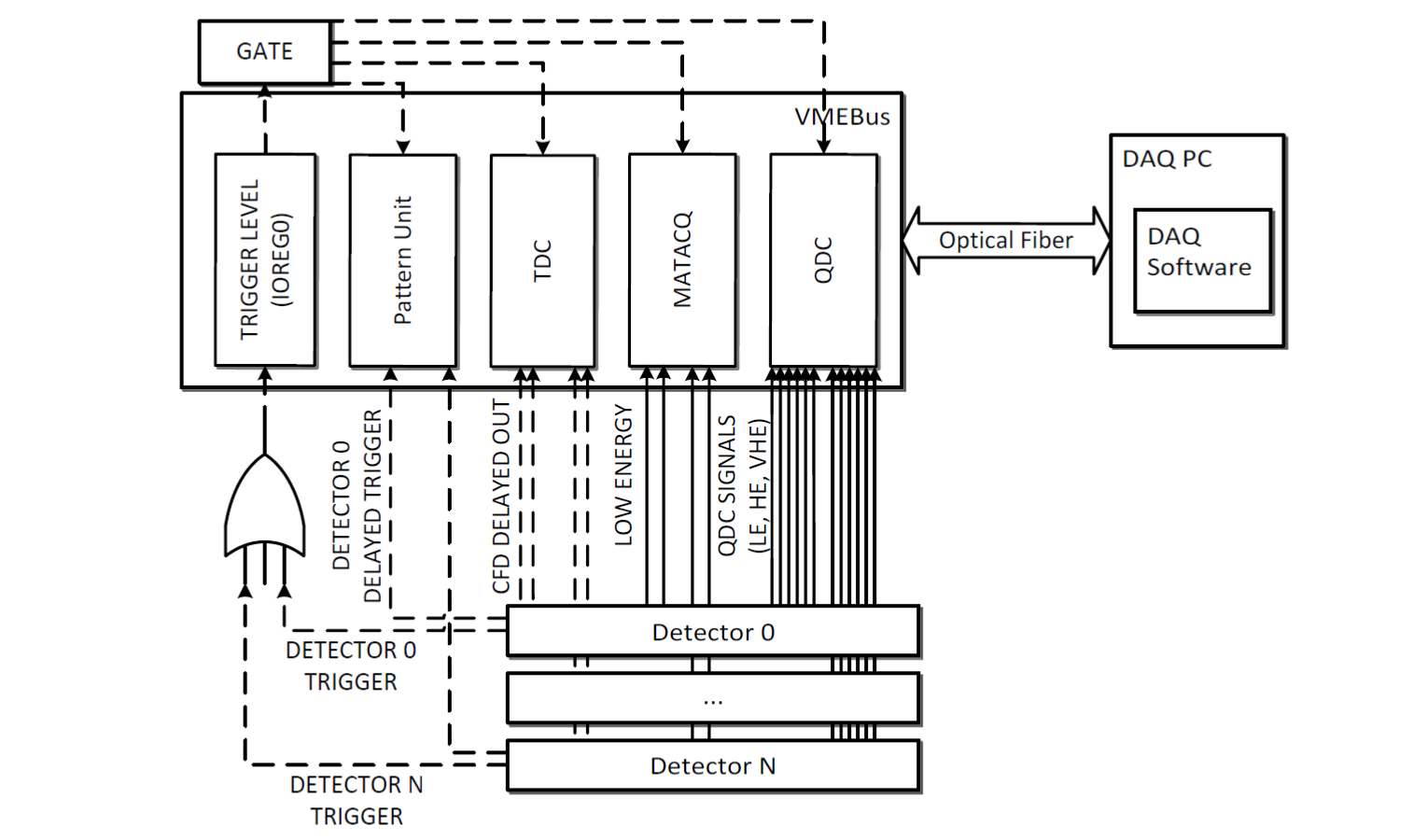}}
\caption{Electronics front-end from one ANAIS module: analog (a) and digital (b) stages.}\label{fig:front-end}
\end{center}
\end{figure}

In this work, full ANAIS acquisition system has been used to get most of the data under analysis. This ANAIS data acquisition system was designed~\cite{TesisMAO} according to the following requirements: trigger at p.e. level in each PMT limiting dark current contribution, high stability, robustness, and scalability. Each PMT charge output signal is separately processed in order to obtain the trigger, pulse shape digitization and energy at different ranges. The two main stages in the acquisition system and some relevant parameters are highlighted below (see Fig.~\ref{fig:front-end}): 
\begin{itemize}
\item Analog stage: PMT signals processing in order to optimize the signal/noise ratio and to produce the acquisition trigger. It consists of a preamplifier, a fan-in/fan-out to produce several identical outputs for a given input, and a constant fraction discriminator (CFD) to produce the trigger for those signals above the stablished threshold at a given fraction of the signal maximum. In a second step, the detector trigger is the logical AND of the two CFD triggers corresponding to the same module  in a 200~ns coincidence window. The global trigger is the logical OR of all the detector triggers. Trigger at p.e. level for each PMT signal and with a good signal/noise ratio has been achieved, and a very high efficiency for the triggering of NaI(Tl) scintillation events at 1~keV level can be confirmed, checked with energy depositions at very low energy (0.87 and 3.2~keV) in coincidence with high energy gammas in a second module coming from the decay of $^{22}$Na and $^{40}$K contaminations in the crystal bulk~\cite{TesisMAO}.
\item Digital stage: Several parameters are saved for every triggered event like PMT signal waveform, charge collected at PMT output, pattern of modules triggering, and time of triggering (real time and accumulated live time). It is based on VME modules which are controlled by the trigger of the analog stage and stores digital information in their buffers. The DAQ program reads these buffers and confims their validity. PMT signal waveform is digitized with a MATACQ-CAEN V1729/V1729A module (12 upgraded to 14 bits resolution and 2~Gs/s sampling for a pulse depth of 2520 points, i.e. a 1260~ns digitization window with 300~ns of pretrigger). Charge-to-Digital converters use a signal integration window of 1~$\mu$s. 
\end{itemize}

\section{Data Analysis}
\label{analysis}
To derive the total light collected, the SER charge distribution for every PMT has to be measured, and then, compared to the charge response for a well-known deposited energy in the scintillator volume. Procedure to determine SER charge distribution is described in subsection~{\ref{SER}}, and is done using the same data that for determining the charge response to a calibration or background line. This allows avoiding many possible systematics in the derived values for the light collected.

\subsection{SER charge distribution determination}
\label{SER}
The SER of the PMTs was measured at the University of Zaragoza, in a dedicated test bench using an UV LED illumination of very low intensity and triggering with the LED excitation signal \cite{TesisMAO}. Afterwards, SER was measured again at LSC, in the same conditions, and even with the same data used for the determination of the charge response corresponding to a known energy deposition. We will refer in the following to the former as {\em test-bench} SER and to the latter as {\em onsite} SER. Comparison of both SER functions allowed to discard possible systematics in the procedure affecting our results.

\begin{figure}[htb]
\begin{center}
\subfigure[]{\includegraphics[width=0.49\textwidth]{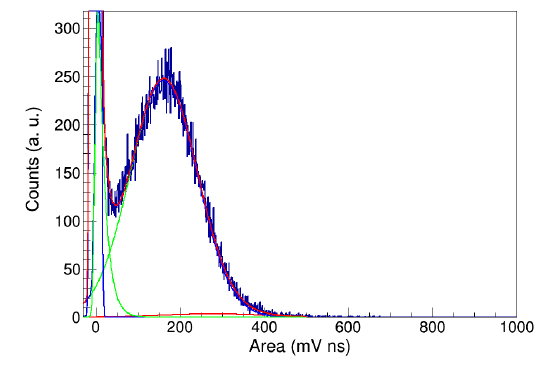}}
\subfigure[]{\includegraphics[width=0.49\textwidth]{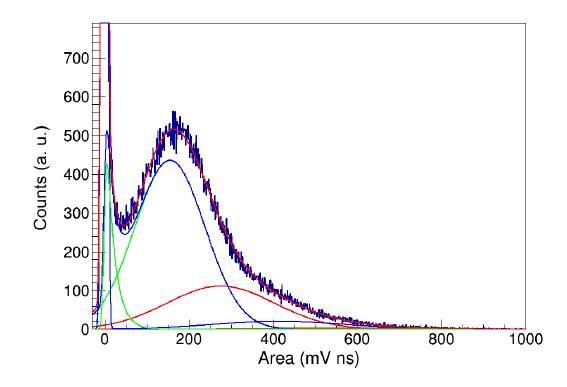}}
\subfigure[]{\includegraphics[width=0.5\textwidth]{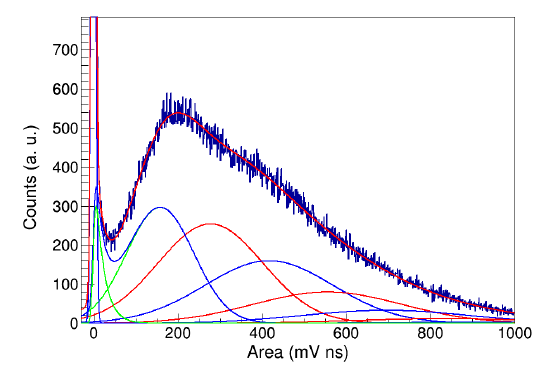}}
  \caption{SER fits for test-bench data corresponding to different illumination conditions: light intensity at PMT is increasing (from (a) to (c))in order to show the contribution of more than one p.e. In (a) and (b) it dominates the single p.e. contribution, SER, (shown in blue), more p.e. contributions are shown alternating red (even numbers of p.e.) and blue (odd numbers of p.e.).}\label{fig:SERtestbench}
\end{center}
\end{figure}

The {\em test-bench} SER was built with the signal produced by the UV LED used together with a high density filter, which resulted in a very low light intensity reaching the PMT; the system was tuned up to have an average number of p.e. per pulse in the PMT much smaller than 1. In that regime, contribution from higher number of incident photons is not observed, and the corresponding charge distribution can be fitted to a gaussian baseline noise, plus a SER function consisting of an exponential component plus a gaussian component. In Fig.~\ref{fig:SERtestbench} it can be seen such a fit for one of the tested PMTs and for different light intensities reaching the PMT in order to enhance the contribution of several p.e. Results of all the fits produced compatible values of the SER parameters \cite{TesisMAO}. 

\begin{figure}[htb]
\begin{center}
 \includegraphics[width=0.8\textwidth]{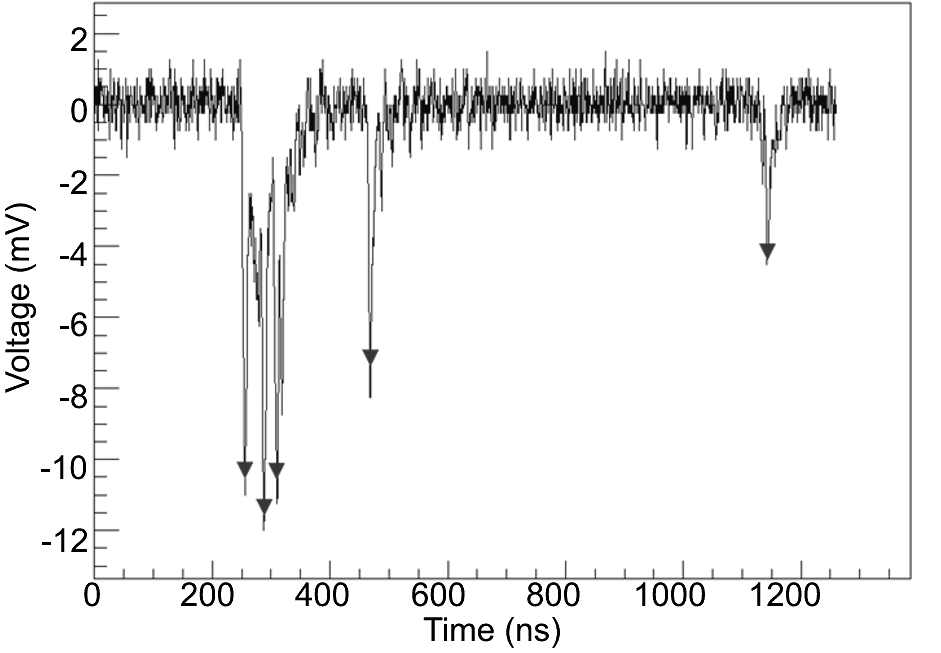}
  \caption{Pulse having a low number of p.e. Peaks identified by the applied algorithm are shown with triangles. The last peak of pulses similar to this are used to build the {\em onsite} SER.}\label{fig:pulse}
\end{center}
\end{figure}

The {\em onsite} SER was built from background data taken at LSC along the normal operation of the detectors in ultra-low background conditions. It profited from the peak identification algorithm developed for the ANAIS data analysis~\cite{TesisMAO}: the SER is built with peaks at the end of pulses having a very low number of peaks (one of them is shown in Fig.~\ref{fig:pulse}) in order to avoid trigger bias, and prevent from p.e. pile-up. Once identified the last peak in one of the pulses having a low number of peaks, the SER charge distribution is derived from the peak area (proportional to charge) integrated in a fixed time window around the peak maximum (from 30~ns before until 60~ns after the maximum).

\begin{figure}[htb]
\begin{center}
 \includegraphics[width=0.7\textwidth]{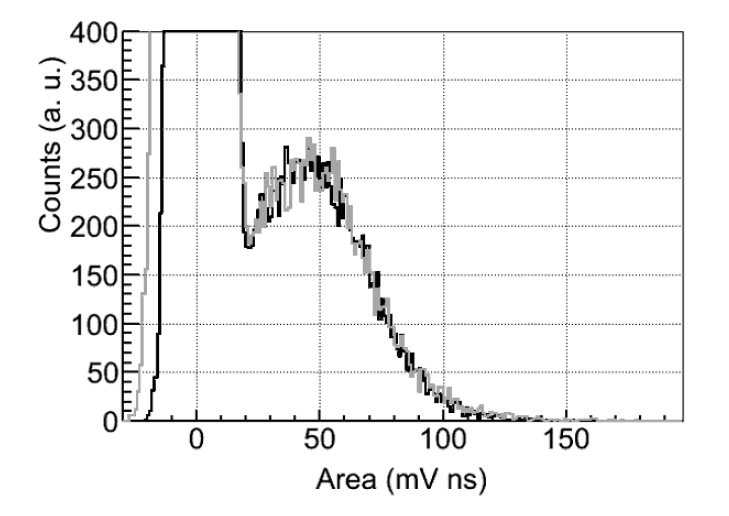}
  \caption{SER charge distributions for the same PMT unit and operation conditions: {\em test bench} SER (gray) and {\em onsite} SER (black).}\label{fig:SERcomp}
\end{center}
\end{figure}

In Fig.~\ref{fig:SERcomp} both SER charge distribution for the same PMT, operated at the same HV bias, are compared. Full agreement validates the use of the {\em onsite} SER in the rest of the paper.

\subsection{Charge response to a known energy deposition}
\label{calresponse}

We have considered for the analysis two different energy depositions at very low energy in the NaI(Tl) crystal: the 22.6~keV x-ray line\footnote{Average value of the different x-ray lines emitted by the source and not resolved by our detectors.} from a $^{109}$Cd external source, and the 3.2~keV line from the decay of $^{40}$K internal contamination. The latter, corresponding to the binding energy of Ar K-shell electrons, has been selected by the coincidence with the high energy gamma produced after the EC to an excited state at 1460.8~keV and detected in a second module \cite{40K}. In the case of the 22.6~keV line, interaction, and then, energy deposition is quite localized, being produced near the source position in the middle of the crystal and near the crystal surface given the mean free path of gammas at that energy in sodium iodide\footnote{NIST Standard Reference Data Program, X-ray attenuation databases.}, 1.8~mm; on the other hand, the 3.2~keV events are distributed throughout the whole NaI(Tl) crystal volume, as they originate in a bulk contamination, homogeneously distributed. Even if the measured potassium content in ANAIS crystals is very low (from 20 to 40~ppb in AS detectors), a few months long measurement allows the identification of the 3.2~keV line; because of the long data taking required for the analysis, 3.2~keV line has not been available in all the considered set-ups (see below). Fig.~\ref{fig:cd} and \ref{fig:40K} show two typical spectra corresponding to a $^{109}$Cd calibration run and to $^{40}$K events selected by the coincidence with the gamma at 1460.8~keV.

\begin{figure}[htb]
\begin{center}
  \includegraphics[width=0.7\textwidth]{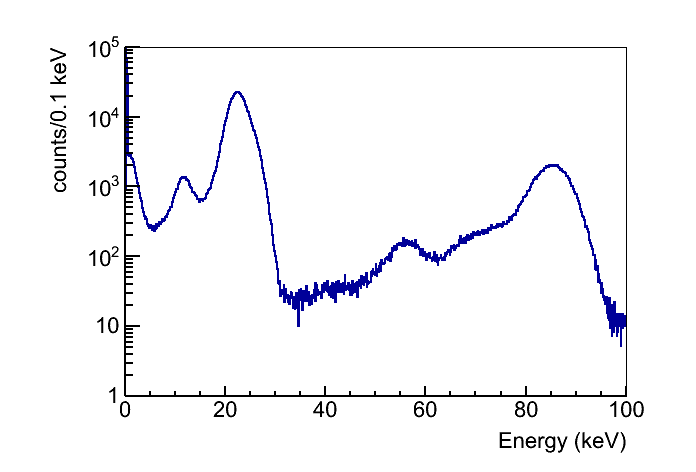}
  \caption{Calibration spectrum corresponding to an external $^{109}$Cd source illuminating D1 module. It features clear lines at 88.0 keV, 22.6 keV and 11.9 keV (the latter corresponds to Br K-shell x-rays from the source plastic cover).}\label{fig:cd}
\end{center}
\end{figure}

\begin{figure}[htb]
\begin{center}
  \includegraphics[width=0.7\textwidth]{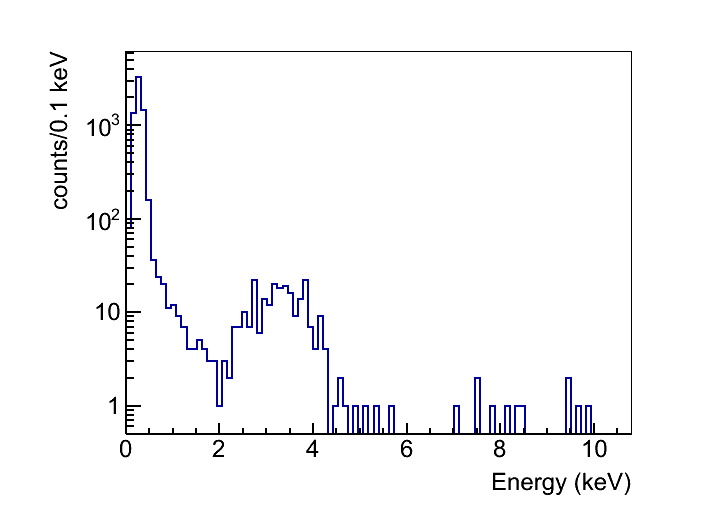}\\
  \caption{Background spectrum from $^{40}$K contamination, built by selecting events at low energy in D1 module in coincidence with a 1460.8~keV deposition in a second module. Statistics is much lower than for calibration lines, as those shown in Fig.~\ref{fig:cd}.}\label{fig:40K}
\end{center}
\end{figure}

\section{Results and conclusions}
\label{results}

Light output per unit of energy deposited in the NaI(Tl) crystal is calculated for each PMT by comparing the mean area associated to a known energy deposition and the mean value of the SER area distribution. Results of this analysis are summarized in Tables~{\ref{tab:phekeV}} and {\ref{tab:lightyield}}: the directly determined number of p.e. detected per keV of deposited energy in each PMT separately, named PMT light output (Table~{\ref{tab:phekeV}}), and the incident light at each PMT and in total required to produce such an output, according to the manufacturer PMT QE nominal values (Table~{\ref{tab:lightyield}}). Uncertainties shown in the tables have been derived by propagating the uncertainties in the calculation of the average SER charge and the average charge for a given energy deposition (either 3.2 or 22.6~keV), both obtained from fits to gaussian functions.

These results are very similar in all the AS modules, while they are clearly better than those obtained for other crystals providers, pointing at a clear manufacturing effect in the optical quality of the system, as far as size of the crystals does not justify those differences. Sharing of the light among the two PMTs coupled to each module, except in the PIII module, is also quite symmetric, as it would be expected, after correcting by corresponding PMT QE nominal values. Asymmetric light output in PIII module PMTs is probably hinting at problems with the optical coupling.

\begin{table}
\begin{center}
\begin{tabular}{| c | c | c | c | c |}
\hline
Crystal & Set-up &  PMT  & \multicolumn{2}{|c|}{Light Output (p.e./keV/PMT)} \\
        &               &        & at 3.2 keV   & at 22.6 keV \\
\hline
\hline
PIII &    A0-5  &  BA0086      &   1.12 $\pm$ 0.21      &          -            \\
     &               & BA0057      &   2.44 $\pm$ 0.20       & - \\
\hline
A0  &    A0-5   & ZK5902  & 3.04 $\pm$ 0.07 & 3.66 $\pm$ 0.02 \\
          &     &  ZK5908  & 3.06 $\pm$ 0.07 &  3.71 $\pm$ 0.07\\
  &   A0-4   & BA0086  & -  & 2.68 $\pm$ 0.04\\
          &    &  BA0057   & - & 2.66 $\pm$ 0.03 \\
  &   A0-3   & BA0086  & - & 2.24 $\pm$ 0.03\\
          &    &  BA0057   & - & 1.78 $\pm$ 0.02 \\

\hline
D0 &   ANAIS25-I  & ZK5902  & -  &  7.77 $\pm$ 0.04  \\
 &   &  ZK5908 & -  &  8.36 $\pm$ 0.66 \\
  &   ANAIS25-III  & ZK5902  & 7.36 $\pm$ 0.39  &  7.64 $\pm$ 0.08  \\
 &   &  ZK5908 & 6.70 $\pm$ 0.21  &  7.61 $\pm$ 0.05 \\
 &   ANAIS37  & ZK5902  & 6.89  $\pm$ 0.19  &  7.38 $\pm$ 0.04  \\
 &   &  ZK5908 & 7.56  $\pm$ 0.22  &  7.88 $\pm$ 0.09 \\
 \hline
 D1 & ANAIS25-I & BA0086  & -  &  5.82 $\pm$ 0.08  \\
 &  & BA0057  & - & 6.76 $\pm$ 0.1 \\
  & ANAIS25-III & FA0018  & 6.54  $\pm$ 0.16  &  7.67 $\pm$ 0.05  \\
 &  & FA0060  & 7.17  $\pm$ 0.25 & 7.52 $\pm$ 0.05 \\
 & ANAIS37 & FA0018  & 6.10  $\pm$ 0.18  &  6.81 $\pm$ 0.05  \\
 &  & FA0060  & 7.73  $\pm$ 0.36 & 7.62 $\pm$ 0.08 \\
\hline
D2  & ANAIS37 & FA0034 & 7.89  $\pm$ 0.44 & 8.21 $\pm$ 0.40\\
  &    &  FA0090 & 7.18  $\pm$ 0.39 & 8.09 $\pm$ 0.38 \\
\hline
D3   & A37D3  &  FA0018  & 6.32 $\pm$ 1.08 &  7.79 $\pm$ 0.08 \\
    &  &  FA0060 & 7.59 $\pm$ 1.27  &  8.15 $\pm$ 0.11 \\
\hline
\end{tabular}
\end{center}
\vspace{0.2cm}
\caption{Light output results for each crystal+PMT in the different configurations tested. Only in the A0-3 set-up light guides were used between the quartz window and the PMT, and then the results are not directly comparable.}
\label{tab:phekeV}
\vspace{0.5cm}
\end{table}

\begin{table}
\begin{center}
\begin{tabular}{| c | c | c | c | c | c | c |}
\hline
Crystal & Set-up &  PMT & \multicolumn{4}{|c|}{Scintillation photons reaching the PMT}\\
&  &  & \multicolumn{2}{|c|}{($\gamma$/keV/PMT)} & \multicolumn{2}{|c|}{ ($\gamma$/keV)}\\
        &               &        & at 3.2 keV   & at 22.6 keV  & at 3.2 keV   & at 22.6 keV\\
\hline
\hline
PIII  &   A0-5    &  BA0086     &    3.9$\pm$0.7      &   -  &     \multirow{2}{1.5cm}{11.3$\pm$0.9}    & \multirow{2}{1cm}{-}        \\
      &                  & BA0057      &   7.4$\pm$0.6     &  - &  &\\
\hline
A0   &  A0-5  & ZK5902  & 8.7$\pm$0.2  & 10.5$\pm$0.1 & \multirow{2}{1.5cm}{17.6$\pm$0.3}  & \multirow{2}{1.5cm}{21.3$\pm$0.2}  \\
   &   & ZK5908  & 8.9$\pm$0.2  &  10.8$\pm$0.2  &   & \\
 &  A0-4 & BA0086  & -  & 9.3$\pm$0.1 & \multirow{2}{1cm}{-}  & \multirow{2}{1.5cm}{17.3$\pm$0.1}  \\
   &   & BA0057  &  - &  8.1$\pm$0.1  &   & \\
  &  A0-3 & BA0086  & -  & 7.8$\pm$0.1 & \multirow{2}{1cm}{-}  & \multirow{2}{1.5cm}{13.2$\pm$0.1}  \\
   &   & BA0057  & -  &  5.4$\pm$0.1  &   & \\

\hline
D0 &   ANAIS25-I  & ZK5902  & -  &  22.3$\pm$0.1 & \multirow{2}{1cm}{-}  & \multirow{2}{1.5cm}{46.6$\pm$1.9}   \\
 &   &  ZK5908 & -  &  24.3$\pm$1.9  &   &  \\
 &   ANAIS25-III  & ZK5902  & 21.1$\pm$1.1  &  21.9$\pm$0.2 & \multirow{2}{1.5cm}{40.6$\pm$1.3}  & \multirow{2}{1.5cm}{44.0$\pm$0.2}   \\
 &   &  ZK5908 & 19.5$\pm$0.6  &  22.1$\pm$0.1  &   &  \\
 &   ANAIS37  & ZK5902  & 19.7$\pm$0.5 &  21.1$\pm$0.1 & \multirow{2}{1.5cm}{41.7$\pm$0.8}  & \multirow{2}{1.5cm}{44.0$\pm$0.3}   \\
 &   &  ZK5908 & 22.0$\pm$0.6 &  22.9$\pm$0.3  &   &  \\
 \hline
 D1 & ANAIS25-I & BA0086  & -  &  20.2$\pm$0.3 & \multirow{2}{1cm}{-}  & \multirow{2}{1.5cm}{40.7$\pm$0.3}  \\
 &  & BA0057  & - & 20.5$\pm$0.1 &  & \\
  & ANAIS25-III & FA0018  & 19.3$\pm$0.5  &  22.6$\pm$0.1 & \multirow{2}{1.5cm}{39.0$\pm$0.9}  & \multirow{2}{1.5cm}{43.3$\pm$0.1}   \\
 &  & FA0060  & 19.7$\pm$0.7  & 20.7$\pm$0.1 &   &\\
 & ANAIS37 & FA0018  & 18.0$\pm$0.5  &  20.1$\pm$0.1 & \multirow{2}{1.5cm}{39.2$\pm$1.1}  & \multirow{2}{1.5cm}{41.0$\pm$0.2}   \\
 &  & FA0060  & 21.2$\pm$1.0 & 20.9$\pm$0.2 & & \\
\hline
D2  & ANAIS37 & FA0034 & 22.0$\pm$1.2 & 22.9$\pm$1.1 & \multirow{2}{1.5cm}{40.4$\pm$1.6}  & \multirow{2}{1.5cm}{43.7$\pm$1.5}  \\
  &    &  FA0090 & 18.4$\pm$1.0 & 20.8$\pm$1.0 &  &  \\
\hline
D3  & A37D3 & FA0018 & 18.6$\pm$3.2 & 23.0$\pm$0.2 & \multirow{2}{1.5cm}{39.5$\pm$4.7}  & \multirow{2}{1.5cm}{45.4$\pm$0.4}  \\
  &    &  FA0060 & 20.9$\pm$3.5 & 22.4$\pm$0.3 &  &  \\
\hline
\end{tabular}
\end{center}
\vspace{0.2cm}
\caption{Scintillation photons reaching each PMT and both for each detector in the different configurations tested, derived from the light outputs reported in Table~{\ref{tab:phekeV}} after correcting by the corresponding PMT QE (Table~{\ref{tab:PMTs}}). Only in the A0-3 set-up light guides were used between the quartz window and the PMT, and then the results are not directly comparable. }
\label{tab:lightyield}
\vspace{0.5cm}
\end{table}

The amount of photons per unit of energy arriving at the PMTs in all the AS modules is outstanding (see Table~{\ref{tab:lightyield}}): more than 40~photons/keV, which is the typically assumed intrinsic scintillation light yield in NaI(Tl) material~\cite{Knoll}. If we consider the well-known nonproportional scintillation response of NaI(Tl)~\cite{JAP}, an increase in light yield at 20~keV of about 20\% with respect to that at 662~keV ($^{137}$Cs gamma line), usually used as calibration standard, can be justified at most, that would mean no more than 50~photons/keV produced by a low energy deposition. Thus, our results would imply a very high number of the photons produced by the interaction reaching the PMT photocathode ($>$80\%) in AS modules. On the other hand, contribution from the slow scintillation components to the total light yield in NaI(Tl) is significant~\cite{NIM_Mos}, and as far as our integration window is 1~$\mu$s, we are not fully including in our estimates all the light produced after the energy deposition. Thus, a relevant conclusion of our measurement is that intrinsic light yield in NaI(Tl), as given in the literature, is probably underestimated in this energy range. Moreover, we observe a systematic lower value for the light incident at the PMTs in every module at 3.2~keV with respect to the value derived at 22.6~keV. This is consistent with the nonproportionality observed in \cite{JAP} at these very low energies, but not with the expected dependence on light yield with Linear Energy Transfer (LET) \cite{Doke,Murray}, which should be much higher for electrons at 3.2~keV, being near the maximum expected scintillation yield. Further efforts to understand scintillation mechanisms in inorganic crystals in the low energy regime are still required. 

The use of light guides 10~cm long between the PMT photocathode and the quartz windows implies a worsening in light collection of about a 25\% at 22.6~keV, by comparing A0-3 vs A0-4 set-ups results for the same crystal and PMTs. This is a very relevant reduction. ANAIS decided to optimize light collection in order to guarantee the lowest energy threshold, and no light guides were considered in the ANAIS detectors design~\cite{CCuestaTesis}.
 
\begin{figure}[h]
\begin{center}
  \includegraphics[width=0.8\textwidth]{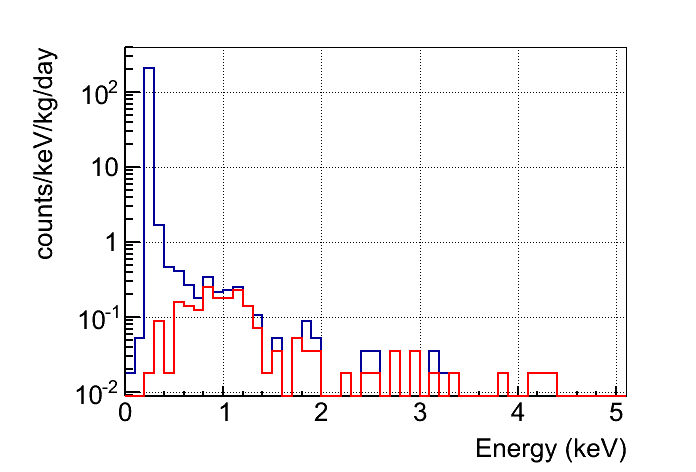}\\
  \caption{Events at very low energy in one AS module produced by $^{22}$Na bulk contamination, selected by the coincidence with a 1274.5~keV energy deposition in a second module are shown in blue; those from these events triggering the acquisition are shown in red. A very high trigger efficiency down to 1~keV can be observed.}\label{fig:0_9keV}
\end{center}
\end{figure}

\begin{figure}[h]
\begin{center}
  \includegraphics[width=0.9\textwidth]{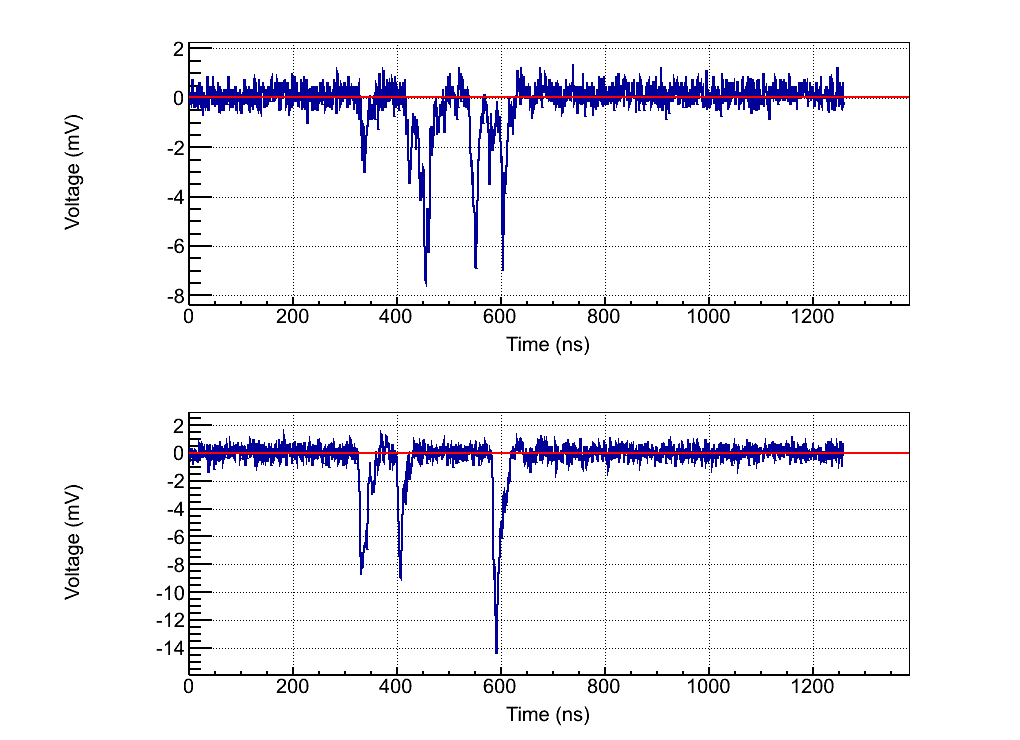}\\
  \caption{One of the events in the 0.87~keV peak shown in Fig.~{\ref{fig:0_9keV}}. Waveforms from both PMTs are shown separately.}\label{fig:pulso0_9keV}
\end{center}
\end{figure}

The reduction of the energy threshold in NaI(Tl) direct dark matter searches down to 1~keV allows a significant improvement in sensitivity to the parameter space region compatible with the annual modulation observed by DAMA/LIBRA experiment~\cite{DAMALIBRA,TesisPV}. The very high light collection measured in ANAIS detectors is for sure very relevant to guarantee such an energy threshold at 1~keV, as far as the mean number of p.e. detected for this energy deposition with two PMTs is around 15 (see Table~\ref{tab:phekeV}). Trigger efficiency is of about 100\% in these conditions, as it has been confirmed in ANAIS analyzing the observed $^{22}$Na bulk contamination cosmogenically induced in the crystals. In Fig.~{\ref{fig:0_9keV}} events selected by the coincidence with a 1274.5~keV gamma in a second detector are shown in blue, while those triggering our acquisition are shown in red for one of the ANAIS AS modules; a clear line at 0.87~keV corresponding to the Ne K-binding energy released after the $^{22}$Na EC decay is observed. In Fig.~{\ref{fig:pulso0_9keV}} one event corresponding to that 0.87~keV line is shown. 

\begin{figure}[h]
\begin{center}
  \includegraphics[width=0.9\textwidth]{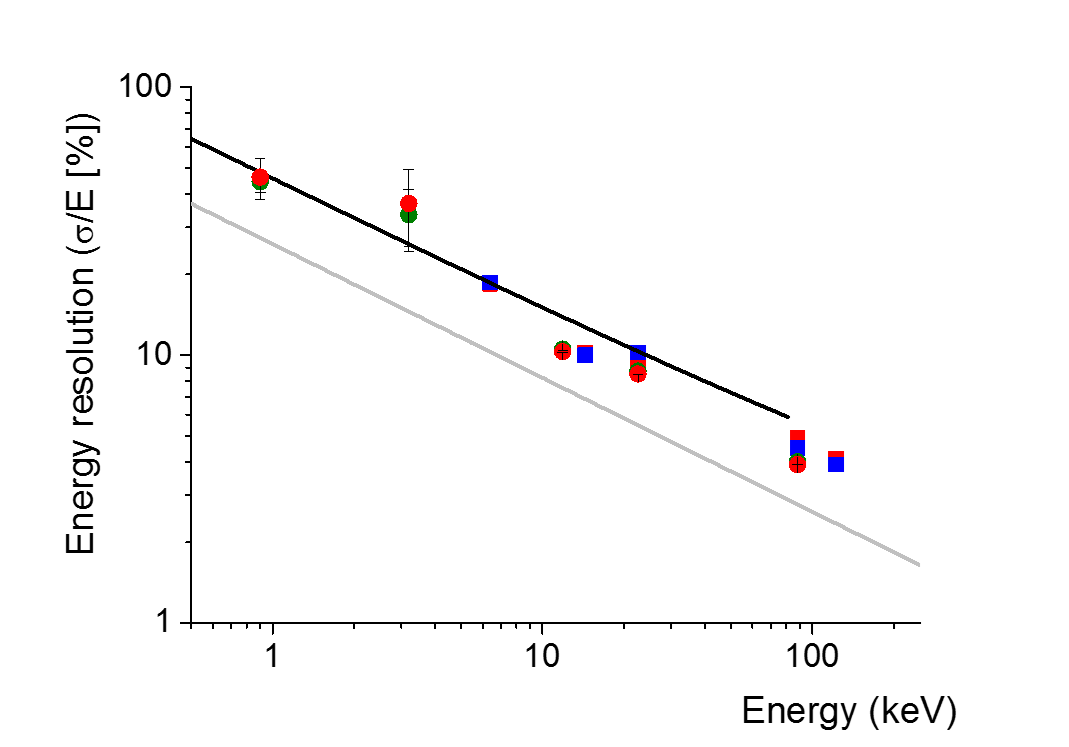}\\
  \caption{Energy resolution (\%) for the different AS detectors below 100~keV (dots and squares). Statistical limit for a light collection of 15~p.e./keV is shown in gray, while in black, DAMA/LIBRA published~\cite{DAMALIBRAapparatus} energy resolution fit at low energy is shown for comparison. Some of the lines are produced using external sources but others are obtained from long background measurements and correspond to bulk energy depositions distributed homogeneously throughout the crystal.}\label{fig:resolution}
\end{center}
\end{figure}

However, the effect in the energy resolution is not so important as it could be thought, because energy resolution in inorganic scintillators is mainly dominated by the nonproportional characteristics of the scintillation~\cite{NIM_Mos}. For instance, total light collection reported by DAMA/LIBRA in phase 1~\cite{DAMALIBRAapparatus} is between 5.5 and 7.5~p.e./keV, a factor of two below the values reported in this paper, but energy resolution is not as different, as can be seen in Fig.~\ref{fig:resolution} where energy resolution is shown for AS modules and DAMA/LIBRA experiment together with statistical limit for a total light collection of 15~p.e./keV. It has to be remarked that some of the lines shown in this plot do not correspond to calibration with external sources, but to long background measurements and bulk energy depositions, which implies that energy resolution could be easily degraded by slow gain drifts and spatial effects on light collection, not affecting data derived from external calibrations.
Notice that some of the energy resolution values shown in Fig.~\ref{fig:resolution} do not correspond to intrinsic energy resolution from monoenergetic lines, but to overall energy resolution in those cases several x-rays contributions are not resolved. At MeV scale, typical values of energy resolution between 2.5 and 3.5\% are obtained, independently of the crystal size, pointing at nonproportionality as dominating effect instead of photon propagation or spatial dependences~\cite{JAP,NIM_Mos}. 

\bigskip

\section*{Acknowledgments}

This work has been financially supported by the Spanish Ministerio de Econom{\'\i}a y Competitividad and the European Regional Development Fund (MINECO-FEDER) under grants No. FPA2011-23749 and FPA2014-55986-P, the Conso-lider-Ingenio 2010 Programme under grants MULTIDARK CSD2009-00064 and CPAN CSD2007-00042 and the Gobierno de Arag{\'o}n and the European Social Fund (Group in Nuclear and Astroparticle Physics). P. Villar was supported by the MINECO Subprograma de Formaci{\'o}n de Personal Investigador. We acknowledge the technical support from LSC and GIFNA staff.

\end{document}